# Intensity Frontier Instrumentation

**Instrumentation Convenors: M. Demarteau, R. Lipton, H. Nicholson, I. Shipsey**
**Editors: S. Kettell, R. Rameika, B. Tschirhart**

The fundamental origin of flavor in the Standard Model (SM) remains a mystery. Despite the roughly eighty years since Rabi asked "Who ordered that?" upon learning of the discovery of the muon, we have not understood the reason that there are three generations or, more recently, why the quark and neutrino mixing matrices and masses are so different. The solution to the flavor problem would give profound insights into physics beyond the Standard Model (BSM) and tell us about the couplings and the mass scale at which the next level of insight can be found.

The SM fails to explain all observed phenomena: new interactions and yet unseen particles must exist. They may manifest themselves by causing SM reactions to differ from often very precise predictions. The Intensity Frontier (1) explores these fundamental questions by searching for new physics in extremely rare processes or those forbidden in the SM. This often requires massive and/or extremely finely tuned detectors.

## Executive Summary

There is overlap of instrumentation needs between kaon and muon experiments. Most of these require detection of particles in the 10 MeV to 1 GeV range associated with stopped kaon and muon interactions. Neutrino experiments tend to have a somewhat broader energy range and somewhat different instrumentation needs.

1) A concentrated R&D effort in the development of large mass, cost effective detectors for neutrino detection and proton decay should be pursued. The most promising technologies for neutrino detection are fully active liquid detectors using water, scintillator or noble liquids. Other possibilities include magnetized iron sampling detectors. All of these types of detectors benefit from more cost effective detection of photoelectrons from scintillation or Cherenkov radiation. Good timing is in general beneficial as well (sub ns). Muon charge identification is a promising option and R&D towards a magnetized LAr detector could be extremely useful, as the LAr detector performance far exceeds that of an iron-scintillator sampling detector. Calibration of these large detectors has a number of challenges and R&D towards calibration systems for these large detectors should be supported.
2) R&D towards cost effective calorimeters with good photon pointing and Time of Flight (TOF) (goal is <20mrad, 10's of psec).
3) R&D towards cost effective, high efficiency photon detection for kaon experiments (with inefficiencies of $10^{-4}$ per photon)
4) R&D to develop very fast, very high resolution photon/electron calorimetry for muon experiments (goal is 100ps, sub-percent energy resolution)
5) Development of very low mass, high resolution, high-speed tracking for muon and kaon experiments (0.001 $X_0$ per space point, 100ps per track)
6) Development of high fidelity simulation of low energy particle interactions. This would include integration and improvement of interfaces of the wide variety of neutrino generator codes with simulation codes. Development of strategies to effectively simulate $>10^{12}$ particle decays & interactions should also be a priority.
7) A concerted effort to develop high throughput, fault tolerant streaming data acquisition systems (goal of TB/second throughput to PB/year data storage)

## Neutrinos and Proton Decay

Neutrinos are the most abundant source of matter in our universe. We believe there are three neutrinos, though some experimental data hints at the possibility of additional sterile neutrinos (2). The discovery of neutrino mass and mixing has led to a renewed desire to determine the absolute mass scale, mass hierarchy and possible existence of CP violation (3). Aside from dark matter, there is no particle with a smaller interaction probability than the neutrino — making detection of neutrinos challenging. The detectors generally need to be very massive, with many different types, specifically designed to uncover the unique signatures of these ghostly particles. The need for massive detectors provides an important synergy with proton decay searches (4).

**Neutrino sources**

Neutrinos from natural sources derive from the Big Bang, supernovae and cosmic rays. From the elusive meV ($<2 \times 10^{-4}$ eV) Big Bang relic neutrinos to the $10^{17-19}$eV GKZ neutrinos (5), ~20 orders of magnitude in energy are spanned, challenging detection techniques. A relic big-bang neutrino experiment would require ~100 grams of weakly-bound atomic tritium, sub-eV energy resolution and μHz background rates above the tritium endpoint (6). Neutrinos are produced abundantly in the explosion of supernovae. The collective flux of neutrinos from supernova explosions throughout the history of the universe is known as diffuse relic supernovae neutrinos. Detection of these isolated neutrinos is most promising in the energy range of a few×10 MeV (7). Though the precise timing of a supernova occurrence cannot be predicted, we expect that one will occur in our galaxy approximately once every ~50 years, with many neutrinos in a few seconds. Cosmic ray interactions in the Earth's atmosphere are a significant source of neutrinos and are the source for the original detection of neutrino oscillations by Super-K (8).

The first neutrino detection was through antineutrino inverse beta decay at the Savannah River nuclear power plant (9). In the intervening decades many neutrino experiments have been carried out at reactors around the world. High-power reactors, typically designed for electric power generation, provide a free source of neutrinos. Radioactive sources of low-energy neutrinos have been proposed for some experiments although the creation of these high-intensity radioactive sources using reactors requires further technical development.

The first accelerator-based beam of neutrinos (1962) enabled the detection of the second or muon neutrino and the 1988 Nobel Prize for Lederman, Steinberger and Schwartz (10). Subsequent neutrino beams enabled the 1972 detection of neutrino neutral current interactions at the Gargamel bubble chamber (11). In these boosted-pion-decay beams, pions are produced by proton irradiation of a thick target, focused forward with a "magnetic horn" and allowed to decay in a long evacuated pipe; the design issues are relatively well understood. While there are challenges to achieve the proton intensities required for superbeams with mega-watt of protons on target, the difficulties seem relatively straightforward to surmount. There are several challenges associated with measuring neutrino oscillations in superbeams experiments. The beam flavor composition is not simple as kaon and muon decays lead to a nonzero flux of $\nu_e$, and a subdominant population of "wrong sign" $\nu_\mu$ survives the pion charge-selection mechanism. Furthermore, the energy dependence of the neutrino flux is not very well characterized and near detectors are necessary to "measure" the different neutrino fluxes. The charged-current scattering cross section for GeV-scale neutrinos is very hard to model. This is subject of ongoing research and is expected to play a central role in next-generation neutrino experiments.

Stopped pion (and muon) neutrino beams have been employed for oscillation studies, cross section measurements and SM tests (12). Recently, there have been proposals to use this type of neutrino source alone or in combination with conventional beams to study CP violation. A high-intensity stopped-pion source employing novel cyclotrons requires significant R&D, but may also have possible applications in industry (13).

Farther-future oscillation neutrino sources could include "neutrino factories" based on stored muon beams and "beta-beams" based on stored radioactive ion beams. The potential of high intensity, tunability and lepton sign selection are highly desirable. The technical challenges, however, are large. Methods for "cooling" the muon require further development. An international design study for a neutrino factory was completed in 2012 (14).

**Neutrino detectors**

Neutrino energies span an enormous range and in some important neutrino experiments, such as $0\nu\beta\beta$, the neutrinos are not even available for detection. However, some neutrino detector themes are common: they rely on large mass, in which ionization charge and/or scintillation or Cherenkov light detection is the primary or secondary detection mechanisms. For reactor electron antineutrinos, large homogeneous liquid scintillators with high light yield have yielded the best performance (KamLAND, Double Chooz, RENO and Daya Bay). For muon neutrino beam detectors it is important to have good event reconstruction to determine the neutrino interaction flavor and type (Charged Current, CC or Neutral Current, NC). These require good particle identification, including good separation of electrons from gammas. Perhaps the easiest neutrino interaction to detect at GeV energies is the $\nu_\mu$ charged current interaction. In a Cherenkov detector the characteristic ring is sharp and easily distinguishable from the fuzzy ring produced by the electron and photon scattering (15). Muons can also be identified in massive iron based, magnetized detectors, a way to detect both the muon and determine its charge (16).

Tau neutrino interactions are difficult to detect. The tau lepton produced in the $\nu_\tau$ charged current interaction has a c$\tau$ of only ~90 microns. Direct detection of tau neutrinos by DONUT (17) and OPERA (18) used nuclear emulsions to provide the necessary tracking resolution. The Super-K detector was also able to provide evidence of $\nu_\tau$ appearance in a statistical manner (19). Further advances in direct detection remain challenging, although liquid argon detectors hold promise.

Because of the tiny neutrino interaction cross section neutrino detectors must typically be very large. Some experiments require only relatively modest scale (kTon or less) detectors, but long-baseline oscillation and astrophysical neutrino experiments require multi-kTon scale detectors. Three detector technologies are most promising for the next generation of multi-kTon detectors. These are liquid detectors: water Cherenkov, liquid argon and liquid scintillator. All three technologies rely on the detection of scintillation or Cherenkov light, whereas LAr detectors also collect ionization electrons to get excellent event reconstruction. A key element of the R&D program would be increased light yield through the development of improved optical attenuation length and increased numbers of detected photoelectrons per dollar. Of these, water and liquid argon are the most suitable for long-baseline accelerator-based oscillation experiments due to better event reconstruction capabilities. Liquid scintillator detectors are most suitable for long-baseline reactor-based oscillation experiments due to high light yield and modest cost.

Water is the most cost effective detector material even though light yields are typically much lower. Only particles above Cherenkov threshold can be detected. Both low energy (~few MeV) and high energy (~GeV) neutrino detection are possible in large water Cherenkov detectors. Particle type can be determined by evaluating the "fuzziness" of a track: electrons and gammas scatter and shower, whereas muons and pions have sharp tracks; the Cherenkov angle can also be of use for particle identification. Low energy event detection may be enhanced with Gd doping for which neutron capture on Gd produces an 8 MeV gamma cascade; this allows tagging of interactions producing neutrons such as inverse beta decay (20).

An enhancement to water Cherenkov detectors can be obtained by dissolution of scintillator in ultrapure water (21). The addition of scintillator has two consequences: ionizing particles below Cherenkov threshold ($\beta < 0.75$) become detectable by their scintillation light and the total light yield increases. Water-based liquid scintillator (WbLS) is a cost-effective solution for future massive detectors with unique capabilities for exploring physics below the Cherenkov threshold, including proton decay via $p \rightarrow K\nu$. These detectors have the benefit of large amounts of scintillation light in combination with Cherenkov light reconstruction. The same water-based detector could also serve as the near detector for Hyper-K or be used for detection of diffuse supernovae neutrinos. Some double beta decay candidates that are chemically hydrophilic are now accessible using WbLS and "tens of tons" scale experiments may become feasible. R&D to increase the number of detected photoelectrons and improve signal timing can improve water Cherenkov and WbLS detector performance.

Phototubes embedded in ice or suspended in water with megaton scale volumes are sensitive to high energy neutrinos and are primarily designed to study astrophysical objects, although there is oscillation sensitivity with atmospheric neutrinos as well. Examples are IceCube, ANTARES and KM3NET. An enhanced detector could conceivably serve as a long-baseline neutrino beam target.

Liquid argon time projection chambers (LArTPC) do not suffer from Cherenkov threshold limitations and in principle extremely high quality particle reconstruction is possible. The ionization charge is drifted with an electric field and collected on readout wire planes; a 3D track can be reconstructed using charge arrival times. Scintillation signals can allow fast timing of signal events and enhance event localization. Very high purity cryogenic argon is required. Track granularity is determined by wire spacing and in principle very fine-grained tracking can be achieved. Particle identification is obtained by measuring ionization along a track. Because of the excellent, full-particle tracking capability of liquid argon, very high-efficiency particle reconstruction allows a smaller LAr detector to match the efficiency of a water detector of ~6 times the mass for neutrino energies above ~GeV. In principle, low energy physics (<100 MeV, e.g. supernova neutrinos) is possible in LAr, assuming adequate triggering.

The largest operating liquid argon detector is ICARUS in the Gran Sasso National Laboratory (22). In the U.S., ArgoNeut (23) took data and the 35-ton LBNE prototype (24) and 170-ton

MicroBooNE (25) detectors will soon. A key R&D area is increasing performance of large LAr detectors. The current program focuses on purification of LAr, measuring materials contamination and measuring fundamental parameters, such as optical attenuation length and electron diffusion. R&D efforts include operation of LArTPCs in test beams, efficient detection of 128nm scintillation photoelectron and reconstruction software development. Additional R&D focuses on TPC design, looking into both 1-phase and 2-phase readout with US effort for massive detectors focusing on 1-phase readout.

Liquid scintillator detectors consist of large volumes in a homogeneous or segmented volume viewed by photomultiplier tubes. There is a long history of successful kTon-scale scintillation detectors, including the segmented Baksan, MACRO, LVD and NOvA detectors, and the monolithic KamLAND and Borexino detectors. Proposed large future detectors include HanoHano, LENA, RENO-50 and JUNO. Light yield in these detectors can be very high, typically 50 times beyond Cherenkov detectors. This enables both low energy thresholds and good energy resolution. However, low energy neutrino detection requires good radioactive purity. Energy reconstruction is based on the number of photoelectrons detected and particle interaction vertices can be reconstructed by timing; to a lesser extent direction and other properties can be reconstructed. Unfortunately because of the isotropy of scintillation light, directionality and tracking capabilities are relatively weak. Nevertheless, some particle reconstruction is possible using photon timing. This kind of detector excels for detection of low energy (tens of MeV) signals, such as reactor, geo and solar neutrinos; furthermore neutrino-less double-beta-decay candidate isotopes can be added. R&D to improve scintillator light yield and attenuation length can have significant impact. R&D to increase the number of detected photoelectrons and improve signal timing has focused on large area photodetectors based on microchannel plates.

It is highly desirable, and in some cases mandatory, to site next-generation detectors in deep underground laboratories. For neutrinoless double-beta decay searches, low rates of cosmic rays and very deep locations are absolutely essential. It is also desirable to site large detectors for long-baseline beams deep underground, as these detectors can then address a much broader range of physics topics. The depth required depends both on physics topic and on detector technology, as the specific nature of the background will vary according to the particular signal. For reference, see the LBNE depth requirement report (26). For these reasons, plans for the next-generation experimental programs focus primarily on deep underground laboratories. Infrastructure at a common site can be shared between different experiments.

The neutrino working group of the Project-X Physics Study undertook to survey neutrino detector requirements of potential Project-X experiments and the capabilities of available technologies, with the goal of identifying high priority areas for R&D (27). The priorities identified by the PXPS neutrino report include:
- New types of liquid scintillator such as LAB and water-based scintillators.
- Improvements in segmentation and readout to build large, economical, room-temperature scintillator detectors that can provide more fine-grained and complete information about neutrino interactions.
- Further work with liquid argon TPC detectors to more fully exploit both ionization and scintillation and efficiently use the wealth of information provided by these detectors.

- Development of ton-scale single phase low-energy threshold liquid argon neutrino detector for coherent scattering measurement
- Various improvements in materials, readout and analysis

## Kaons

Kaon decays have played a pivotal role in shaping the SM. Prominent examples include parity violation, quark mixing, meson-antimeson oscillations, CP violation, suppression of flavor-changing neutral currents (FCNC), discovery of the GIM mechanism and prediction of charm. Kaons continue to have high impact in constraining possible SM extensions.

A key role is played by the FCNC modes mediated by the quark-level processes s→dνν, and in particular the golden modes $K^+ \to \pi^+ \nu\nu$ and $K_L \to \pi^0 \nu\nu$. Because of the peculiar suppression of the SM amplitude (loop level proportional to $V_{us}^5$) which in general is not present in SM extensions, kaon FCNC modes offer a unique window on the flavor structure of such extensions. This argument by itself provides a strong and model independent motivation to study these modes in the LHC era. Rare kaon decays can elucidate the flavor structure of SM extensions, information that is in general not accessible from high-energy colliders. The actual discovery potential depends on the precision of the prediction for these decays in the SM, the level of constraints from other observables and how well we can measure their branching ratios.

The US experimental program pioneered the study of K→πνν and the future program is evolving to include a charged $K^+ \to \pi^+ \nu\nu$ experiment, ORKA (28), making use of the intense Main Injector beam, as a key step towards a neutral $K_L \to \pi^0 \nu\nu$ KOPIO-like experiment (29) using Project-X beams. An extensive kaon program built around the flagship KOPIO experiment is likely to include an upgraded ORKA, along with a potential time reversal violation experiment in the charged mode along with a more general purpose neutral experiment that would measure the $K_L \to \pi^0 l^+ l^-$ modes. The K→πνν experiments require very high $\pi^0$ detection efficiency, with inefficiency well below $10^{-6}$, for $\pi^0$ photons of ~10 MeV to ~1 GeV. In addition, the $K^+ \to \pi^+ \nu\nu$ experiment needs high light collection from plastic scintillator in a ~1.5T magnetic field. The $K_L \to \pi^0 \nu\nu$ experiment needs excellent energy, time and direction measurements of $\pi^0$ photons.

## Muons

Rare muon decays provide exceptional probes of flavor violation beyond the SM. Observation of charged lepton flavor violation (CLFV) is an unambiguous signal of new physics and muons, because they can be made into intense beams, are the most powerful probe. Charged lepton flavor violation experiments at Project X can probe mass scales up to $O(10^4)$ TeV.

The US experimental program naturally starts with the g-2 (30) and Mu2e (31) experiments currently under construction, using intense muon campus beams fed by the Fermilab Booster. Confirmation of the current discrepancy of g-2 with the SM or the observation of Mu2e would require follow-up experiments. A continuation of this program into the Project-X era would likely include muon-to-electron conversion experiments with a range of nuclear stopping targets and next generation experiments to pursue µ→eγ and µ→3e.

## Super Flavor Factories

Current generation heavy flavor factories (b, τ, and c) have severely constrained physics beyond the SM, notably with discovery and measurement of $B_s \rightarrow \mu\mu$ from CMS (32) and LHCb (33) close to the SM prediction and ever tighter limits on $\tau \rightarrow 3\mu$ and $\tau \rightarrow \mu\gamma$ (34). (Belle, LHCb). Evolution to Belle-II (35) and the LHCb (36) upgrade require higher performance lower mass tracking, particle ID, and breakthrough data acquisition performance. The LHCb upgrade represents the first major particle physics experiment with a fully streaming "triggerless" data acquisition system where zero-suppressed data is fully streamed from frontends to high level event processing and filtering (37). Other next generation experiments including Mu2e and ORKA have fully streaming readout design aspirations as well and will benefit greatly from the LHCb experience. Next generation flavor factories also require and drive the development of radiation hard silicon PMT (SiPM) readout technology which is also broadly applicable to high-rate intensity frontier experiments (38).

## Very High Precision Measurements

Experimental progress on the intensity frontier is driven by ever increasing sensitivity to rare processes and every increasing precision on measurement of fundamental parameters such as the muon anomalous magnetic moment (g-2), Electric Dipole Moments (EDMs) (39), and Moller electron-electron scattering (40).

## Calorimetry

Kaon and muon experiments have challenging electromagnetic calorimetry requirements. The Project-X Physics Study (PXPS) Electromagnetic Calorimetry Group investigated a series of muon, kaon and neutron-antineutron oscillation experiments in existing or proposed pre-Project X versions and in several instances examined whether or not the calorimetric techniques employed in these experiments could be extrapolated to produce viable experiments at Project X (41). In so doing, areas of potentially fruitful R&D in calorimetry were identified. The resulting initiatives have both short term experiment-specific goals and longer term generic objectives.

## Tracking

Kaon and muon experiments have challenging tracking requirements. In addition to excellent position resolution, requirements include: extremely high rate capability (potentially up to 1 MHz/mm$^2$), extremely low mass (<< 1% $X_0$) and in some cases good timing resolution (< 1 ns). The PXPS tracking working group surveyed experimental tracking requirements and capabilities of available technologies, with the goal of identifying high priority areas for R&D (42).

## Data Acquisition

Next generation experiments will benefit greatly from advances in networking and processing technologies driven by the IT and telecommunication revolution. Network switching technologies are now woven into "fabrics" with fully programmable point-to-point connectivity between front-end data sources and processing resources. Standard protocols such as Ethernet on standard fabrics can sustain 10-100 GByte/second of data volume from front-ends to processing.

Forced by power density considerations, modern processing is evolving away from high performance single CPU cores toward a large number of integrated low-power CPU cores. Processing is now also embracing the paradigm of computing "fabrics" which define a new Moore's law of performance. Next generation filter processing must be performed in the context of these high density multi-core fabrics with smaller local cache memory for each computing core, which is a significant evolution challenge for event processing software today. Meeting this challenge is highly motivated by potential gains in sensitivity afforded by the high-level processing of all events, for example up to a factor of x10 for complex events in LHCb and at least a factor of x3 for ORKA over the previous generation of kaon experiments.

Evolution to the computing fabric model is necessary in order to sustain Moore's Law advances, and necessary to exploit the advances in current network fabric performance. This is a common challenge shared by both "online" and "offline" computing which further motivates development of streaming data acquisition architectures.

## Computing and Simulation

As noted in the data acquisition discussion, progress toward streaming acquisition architectures reliant on high level filter processing requires very close integration with "offline" computing which effectively becomes another species in a high level processing ecosystem. Next generation intensity frontier experiments can expect to reasonably and economically steward 1-10 Peta-Bytes of data thanks to the pioneering efforts of the LHC experiments which steward 10-100 Peta-Bytes of data per experiment. Despite these enormous gains in data storage capability in the "Big Data" era, streaming data acquisition systems for intensity frontier experiments will require filtering rejection of greater than 1000. Realizing this, where >99.9% of fully reconstructed are rejected forever, will require a robust fault tolerant "self-aware" computing framework and associated applications in order to capture the benefit of fully streaming architectures.

Simulation strategies to model rare processes often rely on factorizing the problem in order to, for example, estimate background processes at the 1 per trillion level ($10^{12}$). While computing systems now exist to provide for simulating in excess of one billion ($10^9$) events, this computing power cannot be exploited without an associated level of integration maturity in the simulation software. The evolution of the GEANT simulation environment is a major success in this regard, yet there are many simulation challenges for future intensity frontier experiments that require further development. Notably a comprehensive simulation of neutrino-nucleus interactions with state-of-the-art generators does not exist within GEANT4 today, requiring researchers to factorize and patch in leading neutrino-interaction generators with GEANT in order to benefit from the high quality particle cascade simulations available in GEANT. Further progress in the treatment of low energy interactions, neutron transport in particular will also be important to high fidelity modeling of rare processes important to intensity frontier research.

## Conclusions

A robust intensity frontier program to fully exploit Project-X beam opportunities requires a robust program of detector R&D. The foremost focus for this R&D program should be upgrades to the neutrino, kaon and muon programs. A core outline of the experimental program has been discussed above, in which upgrades to LBNE, Mu2e and ORKA will compliment a KOPIO-like experiment and possible mueg and Mu3e experiment. Given the tremendous power and versatility of the Project-X beams, we should expect other scientific opportunities will arise and in fact, such opportunities will be facilitated by a robust detector R&D program.

The rough outline of an R&D program that will enable a world leading onshore US intensity frontier program have been presented above. It is important that these thrusts, synergistic with the intensity frontier program, obtain sufficient funding and as the experimental program continues to develop; sufficient R&D funding will be required from the intensity frontier itself.